# Layer-by-Layer Assembly of Efficient Flame Retardant Coatings Based On High Aspect Ratio Graphene Oxide and Chitosan Capable of Preventing Ignition of PU Foam

Lorenza Maddalena [a], Federico Carosio [a]*, Julio Gomez[b], Guido Saracco[a], Alberto Fina[a]

[a] Dipartimento di Scienza Applicata e Tecnologia, Politecnico di Torino, Alessandria Campus,

Viale Teresa Michel 5, 15121 Alessandria, Italy

[b] Avanzare Innovacion Tecnologica S.L, Avda. Lenticares 4-6. Poligono Industrial Lentiscares, 26370 Navarrete

(la Rioja), Spain

*Corresponding author: Tel/Fax: +390131229303/+390131229399;

e-mail address: federico.carosio@polito.it

**Abstract**

The layer-by-layer (LbL) technique is adopted for the construction of multilayers encompassing chitosan and graphene oxide (GO) platelets capable of improving the flame retardant properties open cell PU foams. The LbL assembly follows a linear growth regime as evaluated by infrared spectroscopy and yields a multilayer structure where GO platelets are embedded within a chitosan continuous matrix. 3 and 6 bi-layers (BL) can efficiently coat the complex 3D structure of the foam and substantially improve its flame retardant properties. 3BL only add 10% to the original mass and can suppress the melt dripping during flammability and reduce both the peak of heat release rate by 54% and the total smoke released by 59% in forced combustion tests. Unprecedented among other LbL assemblies employed for FR purposes, the deposition 6BL is capable of slowing down the release of combustible volatile to the limits of non-ignitability thus preventing ignition in half of the specimens during cone calorimetry tests. This has been ascribed to the formation of a protective coating where the thermally stable char produced by chitosan serves as a continuous matrix embedding GO platelets, which control volatile release while mechanically sustaining the PU foam structure.

**Keywords**: Layer-by-Layer, Chitosan, Graphene oxide, Flame retardancy, polyurethane foams

# 1. Introduction

In recent years, the layer-by-layer (LbL) technique has been exploited as one of the most useful surface modification tool able to create functional coating on several substrates [1]. The principles behind LbL were reported for the first time by Iler [2] in 1966, but the first practical applications are documented only thirty years later by the group of Decher [3]. The most used LbL approach is the one that relies on electrostatic interaction occurring during the alternate adsorption of polyanion and polycation on the selected substrate, but many variants based on different interactions (*e.g.* hydrogen bonding) can be employed [4]. The process is quite simple and tunable as it is affected by several parameters like the nature of the employed polyelectrolyte [4], temperature [5], pH [6], the ionic strength [7] and the presence of counterions [8]. By modulating these parameters it is possible to build coatings with controlled thickness (in the 10-1000 nm range) [5, 6, 9] and introduce functionalization which can influence the surface chemistry of the substrate and change, for example, its wettability [10] or macromolecule anchoring properties [11].

In this way, the LbL has been used in a variety of applications ranging from drug delivery [12], gas barrier [13-15] to sensors [16-18]. Recently, this technique has been employed to build-up coatings oriented towards the fire safety and fire protection fields, proving that it is possible to impart flame retardant (FR) properties by the proper selection of the component of the multilayers assembly and the deposition parameters [19-21]. This has been proven to be an efficient approach as polymer flammability is a typical surface property [22]. Indeed, during combustion, the heat radiated from the flame is transmitted through the surface to the bulk of the material. This triggers the thermal degradation of the polymer with the production of combustible volatiles that diffuse through the superficial layer reaching the gas phase and feeding the flame. It is then apparent that by modifying the exchange between the condensed matter and the flame it is possible to control the burning behaviour of a polymer and obtain a FR effect. In addition, recent European regulations have driven the FR research towards environmentally safe alternatives [23]. Indeed, some of the best performing FR additives have been found to be persistent in the environment, ending up in the food chain. This is leading to limitations in use or complete removal of the chemicals which hazard has been proven [24]. In this context, LbL has been presented as a good candidate for satisfying the need for innovative and green FR because, compared with traditional treatments, it has several advantages. The process is carried out in ambient conditions, uses water as solvents and very low concentrations (normally below 1%wt. in water) and the employed solutions can be recycled after use. The application of LbL in the FR field is documented since 2006 [25]. Through years the composition of the LbL coatings has been directed towards different FR actions. Initially, inorganic or hybrid organic-inorganic coatings containing nanoparticles were employed as thermal shields capable of creating, during combustion, an inorganic barrier that protects the substrate and favour the char formation [26, 27]. Lately the coatings have been designed to mimic intumescent systems by incorporating an acidic source, a carbon source and a blowing agent within the assembly [28-30].

FR LbL coatings with hybrid flame retardant mechanism have been deposited on different substrates but the main focus has been on fabrics and open cell poly(urethane) (PU) foams [31]. PU foams represent an important substrate to protect as they are one of the first items to be ignited in fires, quickly leading to flashover events [32]. Different coating compositions have been proposed in order to reduce the fire threat of PU foams. Sodium montmorillonite (MMT) has been employed with chitosan in order to deposit environmental-friendly coatings [21] or in hybrid intumescent compositions with polyphosphates [33]. Both of these systems were able to suppress the melt dripping phenomenon typical of PU foams and achieve a consistent reduction in heat release rates (up to 50%) during cone calorimetry tests in horizontal configuration. The exploitation of lamellar shape nanoparticles has been demonstrated to confer the best FR properties especially when high aspect ratio platelets were employed [34].

However, the majority of the published papers on PU foam protection deals with the use of inorganic nanoparticles like sodium montmorillonite and vermiculite, while limited attention has been directed towards the use of graphene related materials (GRM). This class of materials showed promising FR results when employed in bulk thermoplastic or thermoset polymer nanocomposites [35, 36]. GRM can be successfully exploited in water based LbL assemblies by employing partially oxidised graphene sheets normally referred as graphene oxide (GO) [37]. From a chemical point of view, GO is negatively charged in water due to the presence of oxygenated functionalization obtained by exposing graphene to strong oxidizers, typically sulphuric acid and potassium permanganate [38]. In this manner, it is possible to prepare stable GO suspensions in water or polar organic solvents. A previously reported study employed low concentration GO suspensions for the production of FR LbL coatings where the main constituents were chitosan and alginate, demonstrating the potential of GO in conferring FR characteristics to PU foams [39].

In the present paper, we address the LbL assembly of an efficient FR coating comprising chitosan (CHIT) and Graphene Oxide (GO) for the protection of open cell PU foams. Large GO nanoplatelets (50 ±4 μm) are the main constituents of the assembly. The main aim is the production of an efficient LbL coating capable of delivering strong FR performances with a reduced number of deposition cycles and the obtainment of a deeper insight on the thermal degradation of this GO-based coating.

Chitosan is a biopolymer and is found in nature only in some fungi but it is easily synthesized by the thermochemical deacetylation of chitin [40, 41], which is largely available in nature. The reduction of acetylated units in chitosan ensure the presence of free amino groups that, in acidic conditions, allow its employment as a cationic polyelectrolyte [42]. Within the coating compositions CHIT represents the continuous matrix that holds together GO platelets in a so-defined brick and mortar structure. Upon the application of a flame or radiative heat flux, CHIT may evolve towards the formation of thermally stable aromatic structures that, along with the presence of high aspect ratio GO, will produce a protective coating capable of protecting the PU foam. The LbL growth of this CHIT/GO assembly was monitored with FT-IR spectroscopy and the morphology of the deposited coatings on PU foams was characterized by scanning electron microscopy (SEM). Flammability and forced combustion behaviour of untreated and LbL-treated foams have been investigated by horizontal flammability testing and cone calorimetry, respectively.

A novel approach was applied to investigate the evolution of the coating during combustion by means of infrared and Raman spectroscopies as well as by electron microscopy.

## 2. Experimental Section

*2.1 Materials*

Polyurethane foam (PU) with a density of 24 g/dm$^3$ and thickness of 15 mm was purchased from the local warehouse. PU foam was washed with deionized water and dried in oven at the temperature of 80°C prior to the LbL deposition. Chitosan (CHIT, 75-85% deacetylated), acetic acid, polyacrylic acid (PAA, solution average Mw ~100,000, 35 wt.-% in H$_2$O) and branched poly(ethylene imine) (BPEI, Mw ~25,000 by Laser Scattering, Mn ~10,000 by Gel Permeation Chromatography, as reported in the material datasheet) used in this work were purchased from Sigma-Aldrich (Milano, Italy). Graphene oxide (GO) as 1% wt suspension in water was supplied by AVANZARE Innovacion Tecnologica (Navarrete -La Rioja, Spain). The detailed description of GO suspension preparation along with Transmission Electron Micrographs (TEM) are reported in the Supplementary material file (Figure S1). Solutions and suspensions employed in this work were prepared using ultrapure water having a resistance of 18.2 MΩ, supplied by a Q20 Millipore system (Milano, Italy). Single side polished (100) silicon wafers were used for monitoring LbL growth.

*2.2 layer-by-layer deposition*

CHIT solution (0.25% wt.) was prepared by adding ultrapure water to the chitosan powder and adjusting pH to 4 with acetic acid. The resulting light yellow solution was kept under magnetic stirring for one night. The GO suspension (1%wt) was diluted with ultrapure water to 0.5% wt. and kept under magnetic stirring for 4 hours. PAA 1% wt. solution was obtained by diluting the original PAA solution with ultrapure water. BPEI was employed at 0.1% wt.

Si wafer was employed as model substrate in order to monitor the LbL growth by FT-IR spectroscopy. The surface of Si wafers was activated by 10 min dipping in the BPEI followed by 10 minutes in the PAA solution. After these two steps the Si wafer was alternately dipped in the CHIT and GO solution/suspension. The dipping time was set to 10 minutes for the first bi-layer (BL, *i.e.*

one CHIT/GO pair) and then reduced to 1 min for subsequent layers. After each deposition step, the Si wafer was washed by static dipping in ultrapure water for 1 min and subsequently dried using dust- and oil-filtered compressed air prior to FTIR analysis. IR spectra were collected after each deposition step, up to 10 BL.

PU foams were first immerged in the PAA solution for 10 minutes in order to activate the surface and create a negatively charged surface. After this activation step, PU foams were alternatively immerged into the positively (CHIT) and negatively (GO) charged baths and washed with ultrapure water after each deposition. During the deposition and washing steps the PU foams were squeezed several times in order to let the solution/suspension or washing water penetrate inside the foam structure. The dipping times were maintained the same as for Si wafer. The process was repeated until 3 and 6 BL were built on PU foam following the schematic description reported in Figure 1. At the end of the process, LbL-treated foams were dried to constant weight in a ventilated oven at 80°C. The mass gain, evaluated by weighting the samples before and after the LbL deposition, was found to be 10and 13% for 3 and 6 BL, respectively.

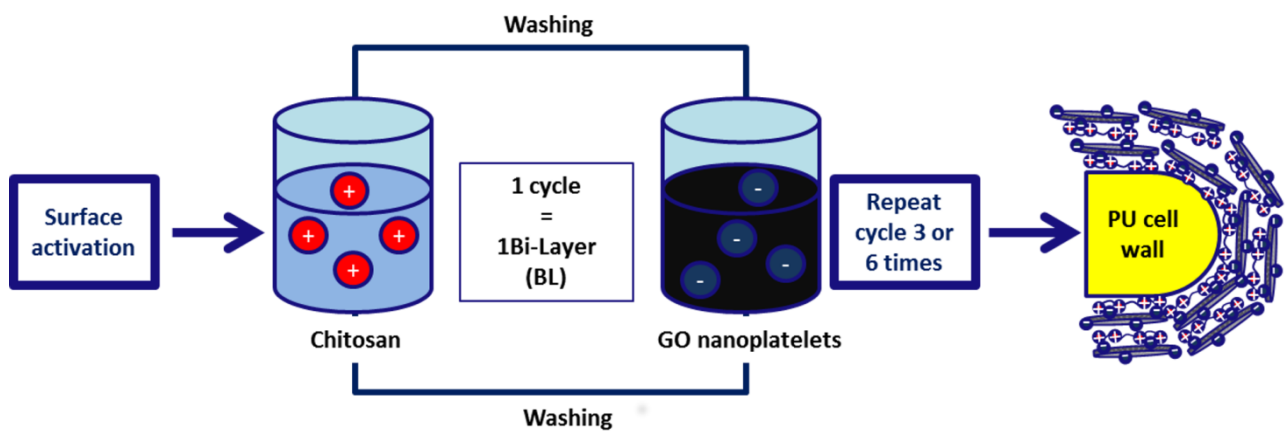

**Figure 1.** Schematic representation of the deposited LbL assembly. PU foams were pre-activated by polyacrylic acid then alternatively dipped in the chitosan solution (positive) and graphene oxide nanoplatelets suspension. The process was repeated in order to deposit 3BL and 6BL.

*2.3 Characterization*

The growth of the layer-by-layer assembly was monitored by FT-IR spectroscopy (Perkin-Elmer Frontier, 32 scansions, 4 cm$^{-1}$ resolution ) using single side polished (100) Si wafer as substrate. The surface morphology of untreated and LbL-treated PU foams was studied using a LEO-1450VP Scanning Electron Microscope (SEM, imaging beam voltage: 5kV). Untreated and LbL-PU foams were cut into small pieces (1 cm$^3$) using a cutter, pinned up on conductive adhesive tapes and gold-metallized prior to SEM imaging. Additional characterization of coatings on Si wafer was performed by high resolution Field Emission Scanning Electron Microscopy (FESEM, Zeiss Merlin 4248, beam voltage: 5kV). Samples were chromium coated prior to FESEM observations.

Flammability tests were performed in horizontal configuration by applying a 20mm methane flame for 3 seconds on the short side of samples (50x150x15 mm$^3$). The test was repeated 3 times for each formulation. During the test, parameters such as final residue and formation of incandescent droplets of molten polymer were evaluated. Prior to flammability tests, all specimens were conditioned at 23±1°C for 48h at 50% R.H. in a climatic chamber. To investigate the combustion behaviours cone calorimetry (Fire Testing Technology, FTT) was employed. 50x50x15 mm$^3$ specimens were analysed under 35kW/m$^2$ radiative heat flux. Measurements were performed four times for each formulation evaluating Time to Ignition (TTI), average and peak of Heat Release Rate (avHRR and pkHRR), Total Heat Release (THR), Total smoke release (TSR) and final residue. Average values and plots are presented. Prior to cone calorimetry tests, all specimens were conditioned at 23±1°C for 48h at 50% R.H. in a climatic chamber. Raman spectra were performed on a InVia Raman Microscope (Renishaw, argon laser source 514 nm/50mW, 10 scans) coupled with a Leica DM 2500 optical microscope. D and G bands were fitted with Lorentzian functions in order to determine their ratio.

## 3. Results and discussion

*3.1 Coating growth by FT-IR spectroscopy*

The LbL growth was monitored using FT-IR spectroscopy. The spectra of neat CHIT and GO were evaluated (Figure S2 and Table S1). The main bands appearing in the spectrum of CHIT are related to the stretching vibration mode of –OH groups in a range between 3700 and 3000 cm$^{-1}$; these signals are overlapped to the stretching vibration mode of amine groups. Alkyl stretching vibrations are visible at 2900 and 2880 cm$^{-1}$ for C-H bond in CH$_2$ and CH$_3$ groups, respectively [43].

However, the most important signals are related to the finger-print region of the spectrum where is possible to identify the asymmetric and symmetric stretching vibration mode of the protonated amine NH$_3^+$ at 1640 and 1556 cm$^{-1}$, respectively [44]. The former signal appears broad due to the presence of the bending vibration mode of water. The peak at 1156 cm$^{-1}$ is attributed to the NH$_3^+$ rocking vibration mode. The most intense band of CHIT is located at 1080 cm$^{-1}$ and it is related to the stretching vibrations of the C-O-C group in the glyosidic linkage [44].

The spectrum of GO evidences the presence of oxygenated species on the surface of graphitic planes with the bands at 1725, 1627 and 1054 cm$^{-1}$ assigned to the stretching vibrations mode of C=O, COO$^-$, and C-O, respectively [45]. Hydroxyl groups stretching vibration modes are also visible in the range between 3700-3000 cm$^{-1}$. The two components have been LbL assembled on Si surfaces following the procedure described in the Materials and methods section. Figure 2 reports 3D projection of restricted IR region, the intensity of the peaks at 1080 and 1624 cm$^{-1}$ plotted as a function of BL number and the FESEM cross section images of the 10BL coating on Si wafer.

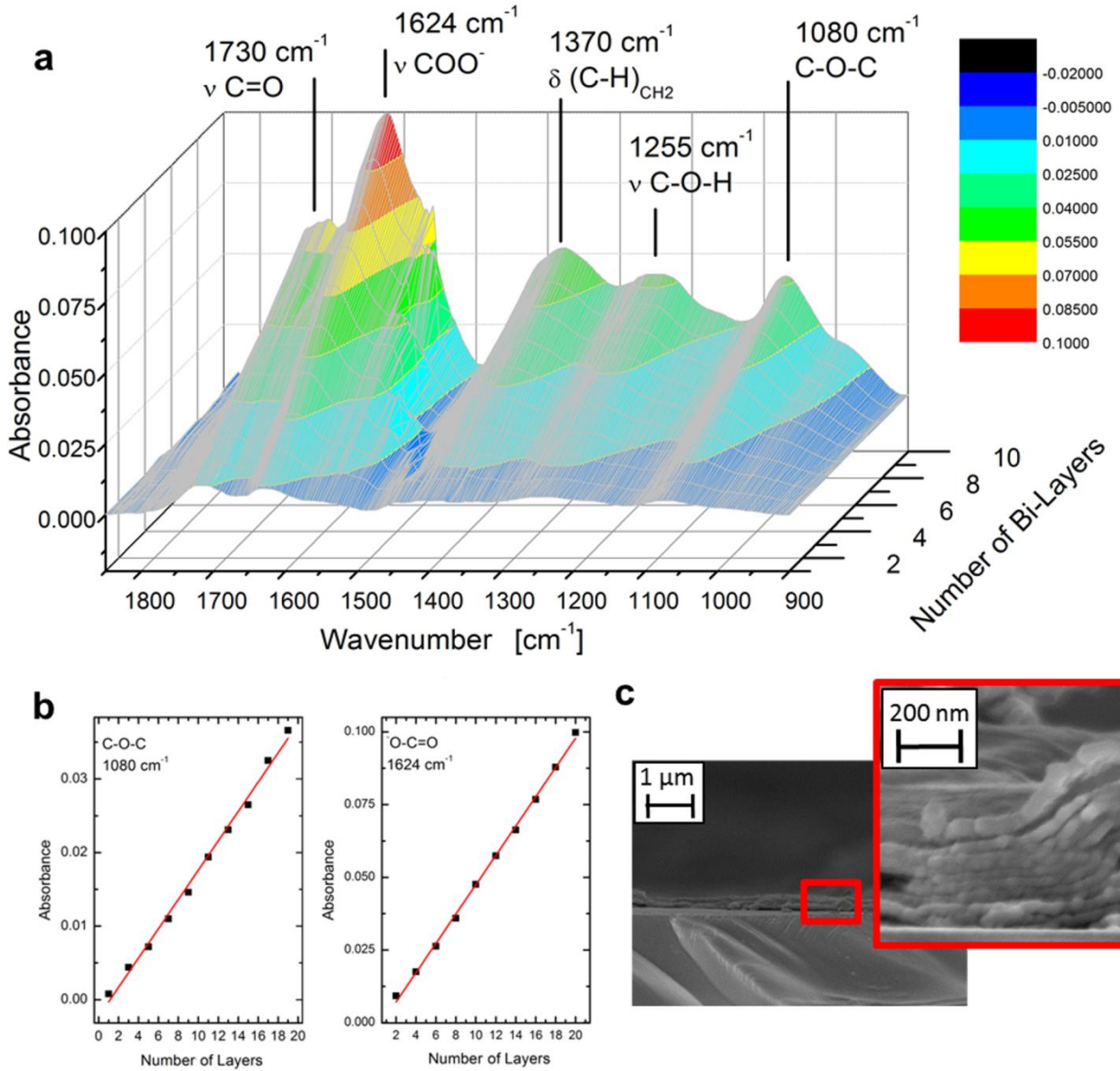

**Figure 2.** a) FT-IR spectra during LbL growth, b) evolution of the signals at 1624 and 1080 cm$^{-1}$ as function of layer number and c) FE-SEM micrographs of 10 BL assembly on Si wafer.

The characteristic peaks of both CHIT and GO can be easily distinguished and grow proportionally to the deposited BL number thus indicating the occurrence of a LbL assembly (Figure 2a). The strongest peak at 1624 cm$^{-1}$ is ascribed to the stretching vibration mode of COO$^-$ in GO and it overlaps with the two CHIT NH$_3^+$ stretching vibrations in the same region (compare Figure S2). Of these two latter, only the symmetric stretching is observable as a shoulder at 1580 cm$^{-1}$. Additional signals, characteristic of both GO and CHIT, can be found at 1730 and 1080 cm$^{-1}$ ascribed to C=O in GO and glyosidic C-O-C in CHIT, respectively. By plotting the absorbance of the signals at 1624 cm$^{-1}$ and at 1080 cm$^{-1}$ as a function of layer number it is apparent that this CHIT/GO system follows a linear growth regime (Figure 2b). This is in agreement with previously reported literature studies dealing with LbL coatings containing CHIT and sodium montmorillonite [21]. The 10 BL cross section on Si wafer has been imaged by FESEM (Figure 2c). The coating appears continuous with some irregularities in thickness due to the wrinkled conformation and high aspect ratio of the deposited GO. High magnification micrographs reveal a layered structure where GO sheets are embedded within a CHIT continuous matrix.

*3.2 Morphology of the coating on PU foams*

Scanning Electron microscopy was employed in order to characterize surface modification after the LbL deposition on PU. Collected micrographs are reported in Figure S3 and Figure 3. Untreated PU foams shows a typical open cell structure (Figure S3); when imaged at high magnifications, the surface of the cell walls is very smooth and homogeneous (Figure 3.a and d). The LbL deposition completely changes this surface morphology without altering the macroscopic structure of the PU foam that maintains its open cell nature (compare Figure S3a, b and c). After the deposition of 3 BL every surface is covered with a continuous nanostructured coating that imparts apparent changes in roughness (Figure 3b and e). This is similar to what already observed on model Si wafer as reported in Figure 2c. By increasing the number of deposited layers the coating increases in thickness and tends to be more rigid leading to the formation of small cracks upon sample cutting prior to SEM imaging (Figure 3c and f). From a mechanical point of view, the coated foams qualitatively appear more rigid while still maintaining the ability of recovering deformation typical of PU foams. This can be ascribed to the presence of the thin and stiff LbL coating that extends through the entire thickness of the foam and thus it is capable of affecting the macroscopic mechanical behaviour of the foam.

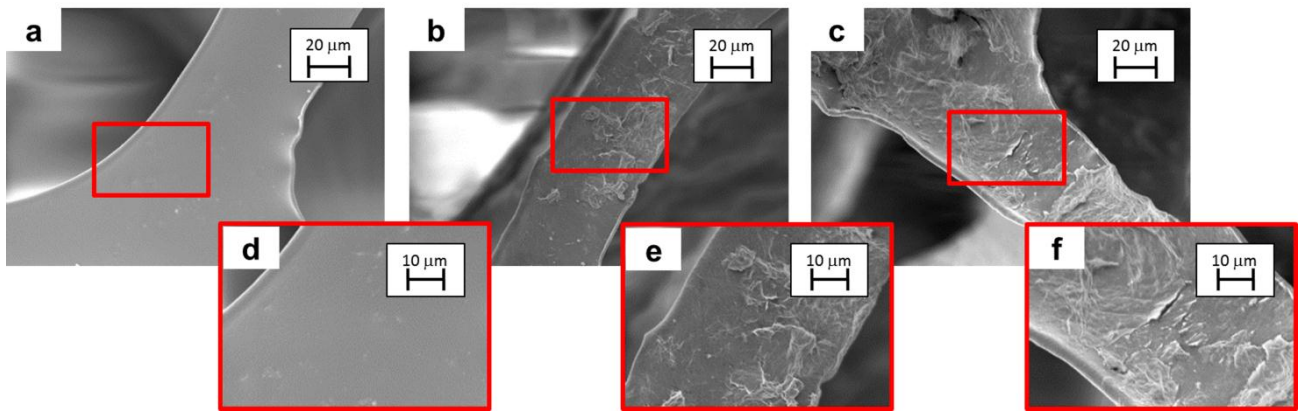

**Figure 3.** SEM micrograph of untreated PU foam (a, d), 3BL PU foam (b, e), and 6 BL PU foam (c, f).

*3.3 Flammability tests*

Because of the large exposed surface and high oxygen permeability, open cell PU are highly flammable. Thus it is important to evaluate the reaction of untreated and LbL-treated foams to a small flame exposure. For this purpose, flammability tests in horizontal configuration were performed: the collected results are summarized in Table 1 while Figure 4 shows digital pictures of the foam during the test.

**Table 1.** Horizontal flammability test data of untreated and LbL treated foams.

| Sample | Dripping and cotton ignition | Residue ±σ [%] |
|---|---|---|
| PU | Yes | 0 |
| 3 BL | No | 61 ± 4 |
| 6 BL | No | 46 ± 3 |

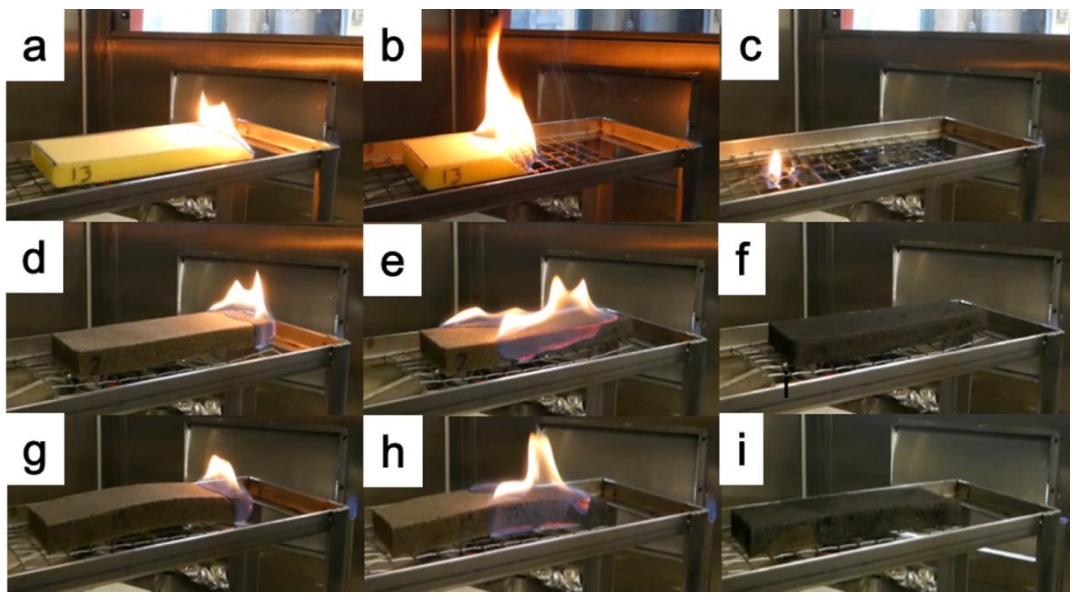

**Figure 4.** Pictures of flammability test in horizontal configuration of untreated PU foam (a-c), 3 BL PU foam (d-f) and 6 BL PU foam (g-i). First column: right after ignition, second column: 15 seconds after ignition and third column: during flame out.

Upon the application of a methane flame the untreated PU ignites instantaneously and burns completely in 63 seconds (see Figure.4 a-c) leaving no residue at the end of the test. During combustion, the formation of the melt dripping phenomenon occurs and flaming droplets of molten PU fall from the sample and ignite the dry cotton placed underneath. This behaviour is well-known for PU foams and represents a potentially dangerous fire threat as it can easily spread the fire to other ignitable materials. The deposition of 3 BL of CHIT/GO significantly changes the burning behaviour of PU foams. Indeed, no melt dripping phenomena occur and, even if the flame is still propagated entirely along the sample by mainly moving on the edges as reported in Figure 4, at the end on the test is possible to collect a self-standing residue that averages 61% of the initial weight (Figure 4.f). In addition, the flames are lower with respect to the untreated PU foam (See Figure 4). The deposition of 6BL yields similar results.

The final residue of the 6BL is lower than the 3BL probably because of the presence of cracks (as observed by SEM images in Figure 3 f) that allow more degradation products from the PU foam to escape and feed the flame. Notwithstanding this, the high residues obtained by both samples point out that the flame self-extinguishes before being able to completely volatilize the PU. From an overall point of view, LbL treated PU foams show an improved behaviour due to the formation of a protective coating that prevents foam collapsing and thermally shields the underlying polymer below the decomposition temperature, thus limiting the release of combustible volatiles.

*3.4 Forced combustion tests*

To better understand combustion behaviour of treated and untreated PU foams, the prepared samples have been tested in forced combustion by cone calorimetry at 35 kW/m$^2$ that correspond to the early stages of a developing fire [46]. Table 2 reports cone calorimetry data of untreated and LbL-treated PU foams. Figure 5 reports HRR and TSR plots as function of time and Figure S4 collects images of the residues at the end of the test.

**Table 2.** Cone calorimetry data of untreated and LbL-treated PU foams.

| Sample | TTI ± σ [s] | Av. HRR ± σ [kW/m$^2$] | pkHRR ± σ [kW/m$^2$] | THR ± σ [MJ/m$^2$] | TSR ± σ [m$^2$/m$^2$] | Residue ± σ [%] |
|---|---|---|---|---|---|---|
| PU | 12 ± 4 | 60 ± 3 | 312 ± 24 | 7.9 ± 0.1 | 78 ± 8 | 6 ± 1 |
| 3 BL | 4 ± 1 | 51 ± 1 | 145 ± 8 | 8.8 ± 0.4 | 32 ± 6 | 7 ± 2 |
| 6 BL | 7 ± 1 | 55 ± 5 | 170 ± 19 | 8.7 ± 0.1 | 58 ± 8 | 6 ± 1 |
| 6 BL* | _ | 20 ± 2 | 42 ± 2 | 3.1 ± 0.5 | 300 ± 14 | 10 ± 1 |

*\* denotes non-igniting samples. During the test only 50% of the samples showed ignition. For non-igniting samples, the behaviour is dominated by thermo-oxidation processes.*

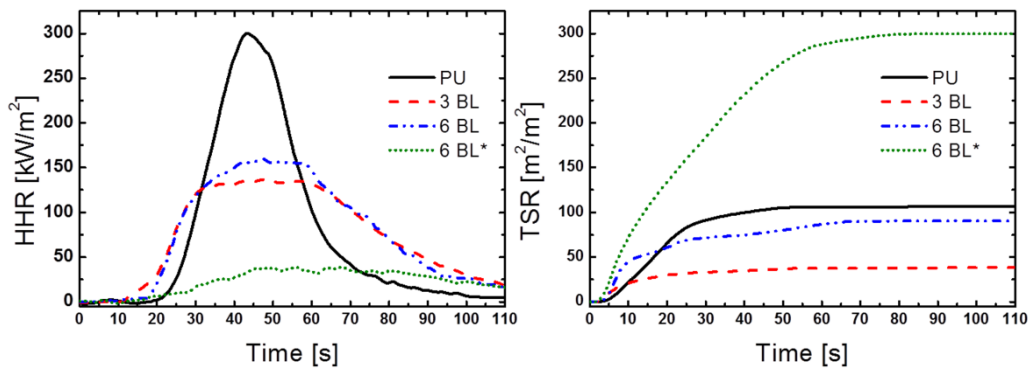

**Figure 5.** a) HRR and b) TSR curves of untreated and PU foams coated with 3 BL and 6BL. 6BL* indicates non-igniting samples.

During the exposure to the cone heat flux, unmodified PU foams quickly melt and partially collapse while releasing volatiles gases that lead to the ignition of the sample. After ignition, the foam completely collapses forming a vigorously burning pool of a low viscosity liquid and reaching the maximum in heat release rate (312 kW/m$^2$) [47]. 3BL samples show a reduction in TTI and a strong reduction in the pkHRR and average HRR (namely − 54 and 15%, respectively). As observed during flammability test, the presence of the LbL coating can prevent the structural collapse of the foam and the high aspect ratio of the employed GO can produce a barrier that slows down volatile release (see digital pictures of the residues in Figure S3). As far as 6BL samples are concerned, two different behaviours have been observed. Indeed, half of the tested samples showed no ignition during test; for this reason, the data and plots of igniting and non-igniting samples have been reported in Table 2 and Figure 5a. Such behaviour can be ascribed to the barrier effect of the coating that slows down volatile release to lower flammability limit. [48] When ignition occurs, the performance of 6BL foams are similar to the 3BL ones. On the other hand, non-igniting samples exhibited very low heat release. Although there is no flame, the sample is subjected to high temperatures so that the PU foam undergoes pyrolysis and a small portion of volatile released is subjected to thermal-oxidation. This process consumes oxygen and is registered by the cone as very low HRR. As far as smoke parameters are concerned, the presence of the CHIT/GO coating can substantially reduce the TSR value as reported in Table 2 and Figure 5b. The highest reduction is achieved for 3BL samples (-59%). On the other hand, as a consequence of the release of degradation products from non-igniting 6BL samples, the measured TSR values are considerably higher than untreated foams although no particulate smoke is actually released. The final residues are not affected by the presence of the coating as they remain within the 6% of the original mass similarly to unmodified PU. This suggests that all the PU is consumed during the test as also confirmed by the increase of THR values reported in Table 2. From cone calorimetry analysis it is apparent that the deposited CHIT/GO multilayers can strongly affect the performance of PU foams in forced combustion tests. Since this test has been widely employed in order to assess the FR properties of LbL coated PU foams it would be of interest to make a comparison with previously published papers. Table 3 summarizes the results in terms of number of layers, mass gain and pkHRR reduction for some most efficient and innovative LbL coatings on PU foams.

**Table 3.** Comparison of pkHRR reduction between the CHIT/GO coatings reported in this paper and previously published LbL coatings on PU foams.

| Composition | N. of monolayers | Coating mass [%] | pkHRR reduction [%] | Ref. |
|---|---|---|---|---|
| **3 BL CHIT-GO** | **6** | **10.2** | **54** | This work |
| **6 BL CHIT-GO** | **12** | **13.4** | **46** | |
| **6 BL CHIT-GO non igniting** | | | **100** | |
| CHIT-GO-Alginate | 30 | 8.3 | 60 | [39] |
| CHIT-MMT pH 3 | 20 | 1.59 | 37 | [21] |
| CHIT-MMT pH 6 | 20 | 4.01 | 52 | [21] |
| PAA-CHIT-PPA-CHIT | 20 | 48 | 55 | [49] |
| Starch3%-SPB23% | 1 | 155 | 75 | [50] |
| Starch3%-SPB5,8%-MMT2% | 1 | 94 | 66 | [50] |

*In the table, PPA: polyphosphoric acid and SPB: sodium polyborates*

All of these coatings were prepared using LbL techniques except for the last two produced by one-pot deposition process that can be still considered as a LbL-inspired approach. The coating composition encompasses natural and synthetic polymers (PAA, CHIT, Alginates, PPA and SPB) and nanoparticles (MMT, GO as in this work). The LbL coatings range from 20 to 30 deposited monolayers, mass add-ons from 8 to 48% and pkHRR reductions in between 37-60%. One-pot deposited coatings exhibit much higher add-on (up to 155%) and the highest pkHRR reduction among the literature works selected for comparison (-75%). From results in Table 3 it is quite apparent that the hybrid CHIT/GO coatings developed within this paper are capable to deliver superior FR performance in terms of pkHRR reduction. In particular, when compared with other assemblies containing GO the coatings developed within this paper can achieve better performances with fewer deposited layers. Indeed, as demonstrated earlier 6 BL can result in non-igniting behaviour with a theoretical 100% reduction in pkHRR. Furthermore, this result is achieved with only 13.4% of added weight. Similarly to what already reported in the literature for inorganic clay containing FR LbL coatings, these impressive results can be ascribed to the high aspect ratio of the employed GO (cfr Figure S1) that allows for better surface coverage and protection [34].

*3.5 Coating evolution during combustion and residue analysis*

In order to better understand the evolution of the deposited LbL-coating during combustion, LbL treated Si wafers have been exposed to the cone heat flux and the changes in coating morphology and chemical composition have been investigated by means of FESEM, IR and Raman spectroscopy. Figure 6 reports collected data and SEM micrographs of the 3 and 6 BL PU foam residues collected at the end of the cone test.

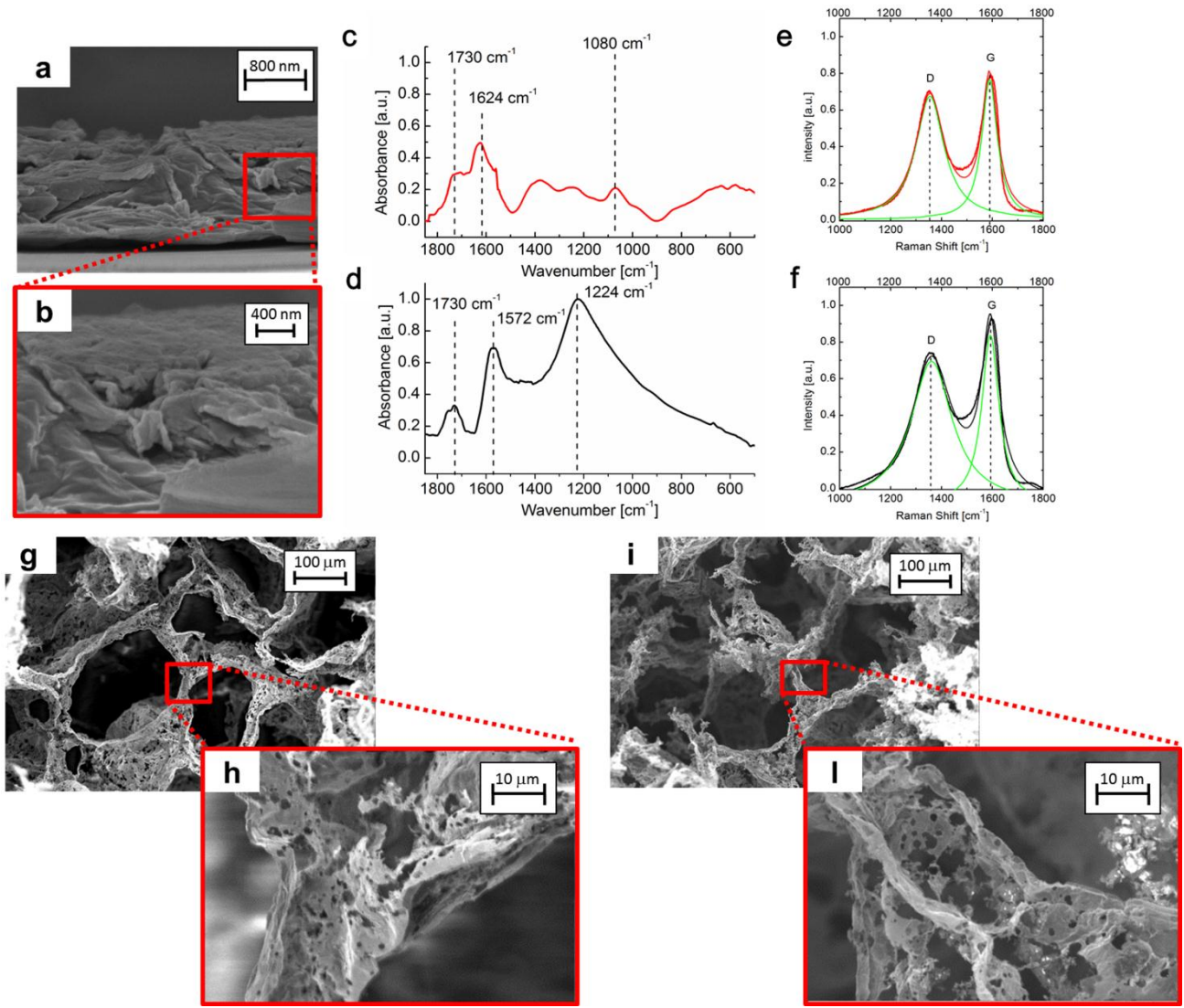

**Figure 6.** FESEM micrograph of 10BL deposited on Si wafer after heat flux exposure (a,b ), FT-IR spectra of 10BL deposited on Si wafer before (c) and after (d) heat flux exposure, Raman spectra of 10BL deposited on Si wafer before (red curve) and after (black curve) heat flux exposure and SEM micrographs of 3 BL (f, g) and 6 BL (h, i) coated foams residues after forced combustion tests.

The cross-section morphology (Figure 6a) reveals a compaction of the coating structure that becomes more irregular with packets of GO platelets more clearly visible with respect to the original morphology (compare Figure 6a with Figure 2c). This behaviour can be ascribed to the thermal degradation of CHIT that, as a consequence of the temperature increase, evolves towards the formation of a carbonaceous char that joins together the GO. This also is confirmed by IR and Raman spectroscopy. Indeed, by IR it is possible to observe a strong change in the signal associated to the LbL coating with the formation of a band at 1572 cm$^{-1}$ characteristic of aromatic carbonaceous structures [43]. This is corroborated by Raman spectroscopy performed on neat CHIT powder before and after cone exposure (Figure S5a) that reveals the formation of two characteristic signals, known as G and D bands, associated to polyaromatic hydrocarbons clearly visible at 1590 and 1350cm$^{-1}$ [51]. Similar bands are present for GO (Figure 5b), in this case the bands are already visible before cone exposure due to the graphene-like structure of GO (Figure S5b). Moreover, by a simple evaluation of the ratio between the area underneath the G and D bands it is possible to obtain information about the quality of the graphitic structure [52]. Indeed, as reported in the literature, the D band is normally associated to defects and is employed to evaluate the quality of graphene-based material. When the two components are LbL assembled the evolution of the Raman spectra show an increase of the D/G ratio (from 1.57 to 1.99) ratio that can be mainly ascribed to the formation of highly disordered char from CHIT.

From the collected characterization it seems that upon heat flux exposure the coating undergoes structural rearrangements due to the thermal degradation of CHIT that produces an aromatic char that glue together the GO in a compacted and thermally stable structure. Being deposited on all available surfaces of the PU foam, this structure can mechanically sustain the foam and prevent its collapse while acting as barrier to volatiles. SEM observations performed on the 3 and 6BL residues collected after cone tests confirm this hypothesis (Figure 6). For both samples the original 3D structure is still visible and high magnification micrographs reveal the presence of a hollow structure mainly made by the charred coating which structure is not continuous and characterized by defects. This confirms the results obtained by flammability and cone calorimetry where it was apparent that although the coating controlled and slowed down volatile release this effect was not enough to allow for early self-extinguishing behavior during flammability or prevent the complete combustion/pyrolysis of the PU during cone calorimetry.

## 4. Conclusions

Multilayered structures comprised by chitosan and high aspect ratio graphene oxide platelets have been assembled using the layer-by-layer technique and employed as flame retardant coating on open cell PU foams. The coating growth has been characterized by infrared spectroscopy pointing out a linear growth regime for this assembly. Electron microscopy observations performed on a 10BL coating cross section revealed a continuous coating characterized by a multi-layered structure where GO platelets are embedded within a continuous chitosan matrix. The same coating also showed an irregular surface due to the wrinkled nature of the employed GO. 3 and 6 BL of this chitosan/GO assembly can conformally coat the complex 3D structure of PU foams yielding a continuous coating that extends through the entire thickness of the foam. By flammability tests in horizontal configuration, LbL treated foams were found able to completely suppress the melt-dripping phenomenon and self-extinguish the flame before the complete consumption of the PU yielding final residues as high as 61%. During cone calorimetry tests the deposition of only 3BL reduced the pkHRR by 54% and TSR by 59%. Surprisingly, 6BL were able to prevent foam ignition on half of tested samples indicating that the coating can slow down the volatile release to lower flammability limit. This behaviour, to the best of the author's knowledge, has never been reported for LbL coated PU foams. The coating evolution upon exposure to the cone calorimeter heat flux has been also investigated. During combustion the chitosan degrades producing a stable aromatic char that glues together the GO platelets in a barrier that can mechanically sustain the foam while slowing down the release of volatiles. The present paper makes it possible for a further step towards safer PU foams and offers a FR solution to be further developed and extended to other substrates.

## Acknowledgments

This work was funded by the Graphene Flagship Core 1, grant agreement n° 696656. Mr. Mauro Raimondo and Mr. Fabio Cuttica are acknowledged for FE-SEM analyses and cone calorimetry tests, respectively. In addition, we thank Dr. Mar Bernal for the help and valuable discussions.

*Author contributions*

F. Carosio and A. Fina conceived the experiments, F. Carosio coordinated the project, L. Maddalena carried the LbL deposition and the characterization. J. Gomez provided the GO suspension and contributed to the discussion. A. Fina and G. Saracco contributed to the discussion and interpretation of the results. The manuscript was mainly written by L. Maddalena and F. Carosio.

*Supplementary material*

Detailed description of graphene oxide preparation, TEM micrographs of graphene oxide, FT-IR of neat chitosan and graphene oxide, low magnification SEM micrographs of untreated and LbL-treated PU foams, digital images of cone calorimeter residues, Raman spectra of neat chitosan and graphene oxide before and after heat flux exposure are supplied as Supplementary material.

# Supplementary material

**Layer-by-Layer Assembly of Efficient Flame Retardant Coatings Based On High Aspect Ratio Graphene Oxide and Chitosan Capable of Preventing Ignition of PU Foam**

Lorenza Maddalena[a], Federico Carosio [a]*, Julio Gomez[b], Guido Saracco[a], Alberto Fina[a]

[a] Dipartimento di Scienza Applicata e Tecnologia, Politecnico di Torino, Alessandria Campus,

Via Teresa Michel 5, 15121 Alessandria, Italy

[b] Avanzare Innovacion Tecnologica S.L, Avda. Lenticares 4-6. Poligono Industrial Lentiscares, 26370 Navarrete

(la Rioja), Spain

*Corresponding author: Tel/Fax: +390131229303/+390131229399;

e-mail address: federico.carosio@polito.it

*Graphene oxide preparation*

Water dispersion of Graphene oxide (GO) was prepared using a modified Hummers' method in $H_2SO_4$. Starting from large flakes of natural graphite (NGS-Naturgraphit) and using a proportion of graphite/$KMnO_4$/$NaNO_3$ 1:3.75:0.25. The reaction temperature inside the reactor was kept between 0 and 4 ºC during the oxidants addition (reaction time 72 h). The resulting solution was slowly warmed up to 20ºC and maintained for 72 hours of reaction. To remove the excess of $MnO_4^-$, $H_2O_2$ solution was added to the reaction mixture and stirred overnight. After sedimentation, the solution was washed with HCl 4 %wt solution by 2 h under mechanical stirring. The solid was filtered obtaining wet graphite oxide. Wet graphite oxide was dispersed in osmotic water (1 % wt based on dry graphene oxide) and stirred in a in a Dispermat LC75 using a cowless helix at 1000 rpms for 10 minutes and them at 20.000 rpms for 60 seconds. This dispersion was ultrasonicated with a UP400S HIELCHER for 1 hour using a H40 sonotrode with 90% of amplitude and full cycle condition to exfoliate the graphite oxide and obtain the graphene oxide water dispersion. The pH of the solution was measured during the first 24h after its preparation with a calibrated pH meter obtaining a value of 1.89±0.06. The viscosity of 1%wt GO water suspension has been determined using a rotational viscosimeter Brookfield EVO Expert R. The measurement was done employing a low-viscosity-adapter tool, due to the low viscosity of the suspension. The container with sample was submerged in a thermostatic bath to ensure constant temperature during the measurement (25ºC). The viscosity was found to be 9.7±0.6 cP. For Transmission Electron Microscopy (TEM) characterization, GO water suspensions were dispersed in isopropyl alcohol and sonicated with in a COBOS bath sonicator for 15 minutes. Transmission electron microscopy (TEM) experiments were performed on a JEOL model JEM-2010 electron microscope. The resulting micrographs are reported in Figure S1. The average lateral size has been evaluated as 50±4 microns based on Lased diffraction particle size distribution.

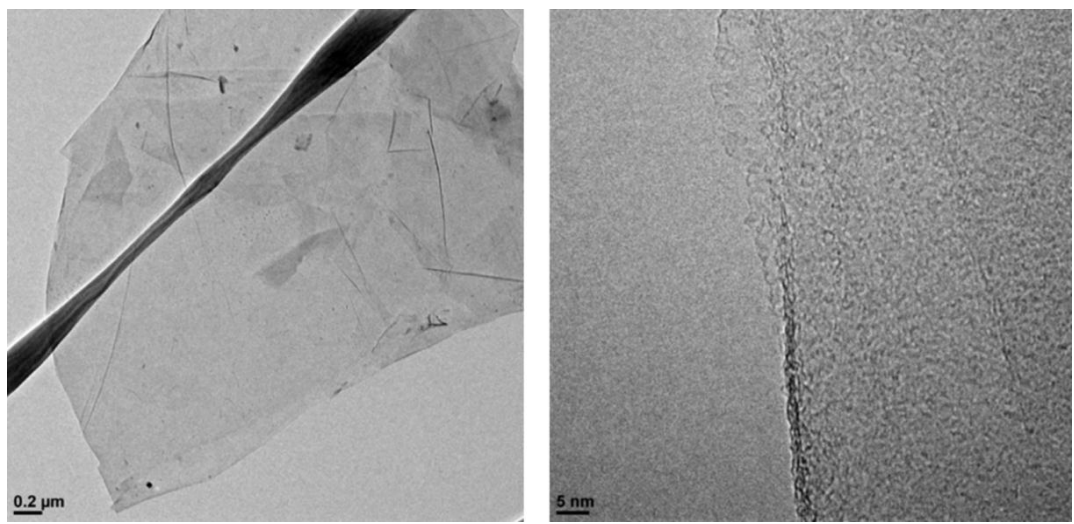

**Figure S1.** TEM micrographs of graphene oxide (GO) at different magnifications

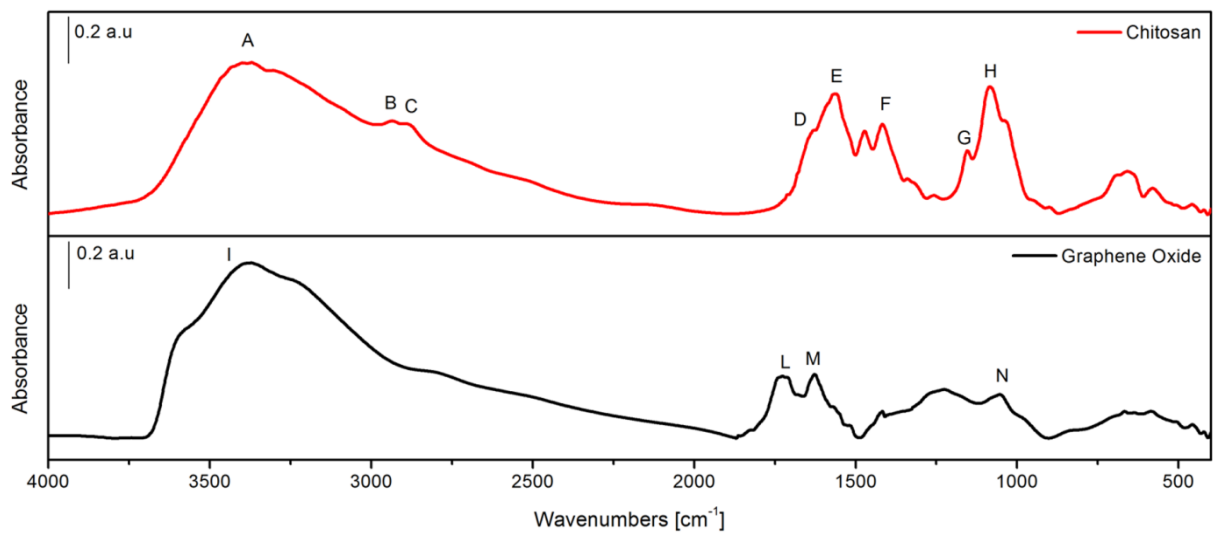

**Figure S2.** FT-IR spectra of pure Chitosan (red curve) and Graphene Oxide (black curve)

**Table S1.** Signals attribution of Chitosan and Graphene Oxide spectra.

| | Signal | Wavenumbers (cm$^{-1}$) | Attribution |
|---|---|---|---|
| | A | 3800 - 3000 | $\nu$ (O-H) and $\nu$ (N-H) |
| | B | 2900 | $\nu$ (C-H) of $CH_2$ |
| | C | 2880 | $\nu$ (C-H) of $CH_3$ |
| **Chitosan** | D | 1640 | $\nu_{as}$ ($NH_3^+$), $\nu$ (O-H)$_{H_2O}$ |
| | E | 1556 | $\nu_s$ ($NH_3^+$) |
| | F | 1156 | $NH_3^+$ rocking |
| | G | 1411 | $\delta$(C-H) of $CH_2$ |
| | H | 1070 | $\nu$(C-O-C) of glycosidic units |
| | I | 3800 - 3000 | $\nu$ (O-H) |
| **Graphene** | L | 1725 | $\nu$ (C=O) |
| **Oxide** | M | 1627 | $\nu$ ($^-$O-C=O) |
| | N | 1054 | $\nu$ (C-O) |

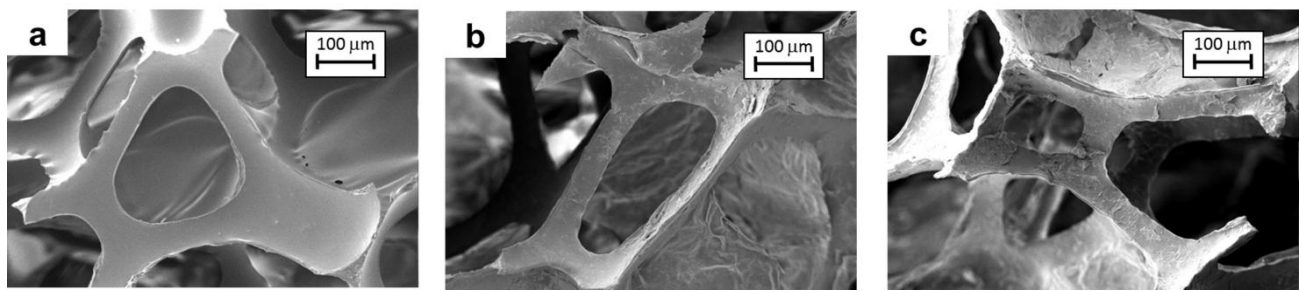

**Figure S3.** SEM micrograph of a) untreated PU foam, b) 3BL and c) 6BL PU foams

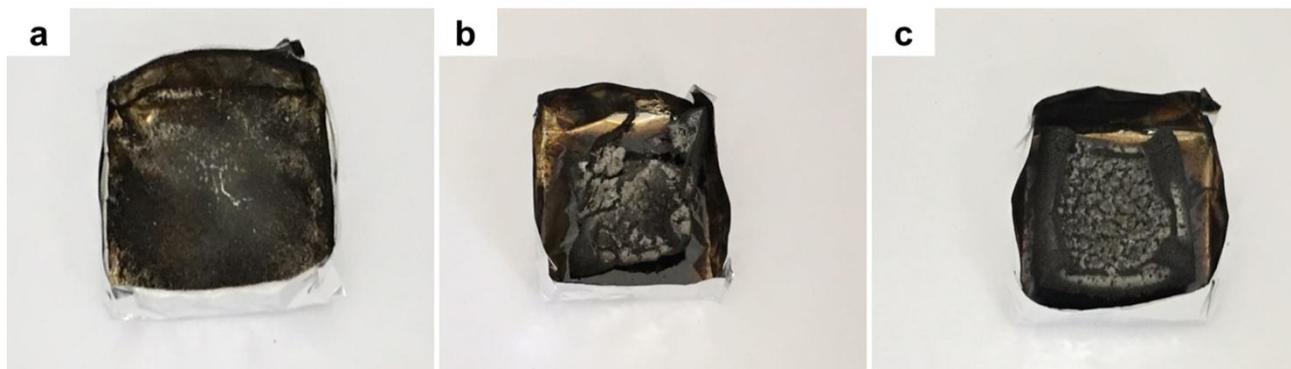

**Figure S4.** Images of a) untreated PU foam, b) 3BL and c) 6BL PU foams cone calorimeter residues

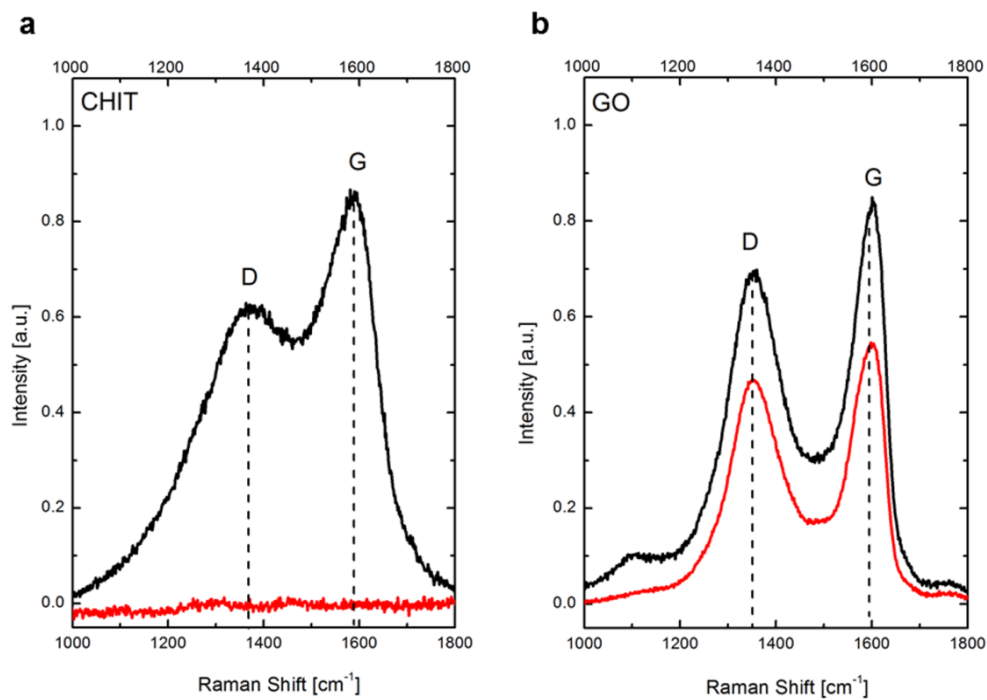

**Figure S5.** Raman spectra of neat GO (a) and after exposure to heat flux (b). Raman spectra of Chit before (red curve) and after (black curve) exposure to heat flux. Table of D and G band ratio (c).